# Validating the Claim "Defeating HaTCh : Building Malicious IP cores"

Anshu Bhardwaj,  Subir Kr Roy
International Institute of Information Technology, Bangalore, India

**Introduction**

Haider et al. in their paper [1] have proposed a rigorous algorithm HaTCh (*Ha*rdware *T*rojan Cat*ch*er) which according to their claim detects any Hardware Trojan (HT) from deterministic Hardware Trojans ($H_D$). They also claim in their paper that $H_D$ constitutes a huge class of deterministic HTs which is orders of magnitude larger than the small subclass (e.g. Trust-Hub) considered in the current literature. Bhardwaj et al. recently published one paper "Defeating HaTCh: Building Malicious IP Cores" [2], which claims that their newly designed hardware Trojan can evade the detection by HaTCh.

In the paper "Comments on "Defeating HaTCh: Building Malicious IP Cores"[3] Haider et al. have countered that HaTCh algorithm is not defeated by the proposed Trojan design of Bhardwaj et al since this trojan is an "always ON" type and hence it does not fall into $H_D$ class of trojan for which HaTCh algorithm is designed.

The efforts put in by designer of HaTCh algorithm for their systematic approach to define four key parameters for Trojan classification in $H_D$ followed by the rigorous analysis carried out by them for justifying the robustness of their algorithm in detecting all trojans in class $H_D$, is commendable. However, their claim that the above-mentioned Trojan cannot defeat HaTCh is incorrect, due to the authors' misunderstanding of implicit trigger mechanism embedded in our proposed Hardware Trojan.

Further, we strongly claim that our proposed Hardware Trojan definitely belongs to the class $H_D$ and that it will definitely not be detected by the HaTCh algorithm. We justify our claim as explained below.

**Properties of $H_D$ Hardware Trojans, *as defined in HaTCh Algorithm[1]***
*(The following text is taken from the paper [1])*

$H_D$ represents the HTs which are embedded in a digital IP core whose output is a function of only its input, and the algorithmic specification of the IP core can exactly predict the IP core behavior. A brief highlight of four crucial properties (d, t, α, l) of this class which determine the stealthiness of a HD HT, is as follows:

  i. *Trigger Signal Dimension d(T)* represents the number of wires used by HT trigger circuitry to activate the payload circuitry in order to exhibit malicious behavior. A large *d* shows a complicated trigger signal, hence it is harder to detect.
  ii. *Payload Propagation Delay t(T)* is the number of cycles required to propagate malicious behaviour to the output port after the HT is triggered. A large *t* means it takes a long time after triggering until the malicious behavior is seen, hence less likely to be detected during testing.

iii. *Implicit Behavior Factor α (T)* represents the probability that given a HT gets triggered, it will not (explicitly) manifest malicious behavior; this behavior is termed as implicit malicious behaviour. Higher probability of implicit malicious behavior means higher stealthiness during testing phase.
iv. *Trigger Signal Locality l(T)* shows the spread of trigger signal wires of the HT across the IP core. Small *l* shows that these wires are in the close vicinity of each other. Large *l* means that these wires are spread out in the circuit, and hence the HT is harder to detect.

A Hardware Trojan can be represented by multiple sets of trigger states T, each having their own d, t, α and l values. The collection of corresponding quadruples (d, t, α, l) is defined as the achievable region of the Hardware Trojan. This allows us to define H $_{d,t,α,l}$ as follows:

*Definition:* H $_{d,t,α,l}$ is defined as all H$_D$ type Trojans which can be represented by a set of trigger states T with parameters d(T)≤d, t(T)≤t, α(T)≤α and l(T)≤ l.

**Brief Description of Trojan described in "Defeating HaTCh…" [2]**

HT design proposed by Bhardwaj et al [2] belongs to the information leakage hardware trojan category that leverages system specifications targeted to provide protection from traffic analysis. The objective of hardware Trojan inserted inside this IP block is to leak secret encryption key over the channel to enable the adversary to decipher the encrypted communication over the channel. It employs self-synchronizing scramblers [4] (also called multiplicative scramblers) to transmit and receive key bits clandestinely, over the same output pin without requiring any fixed known bit pattern to spread the key bits. At the transmitter, the key bit is xored with the feedback bit of LFSR; based on the key bit being '1' or '0' the PRSG output will either be inverted or continue to be in the non-inverted form. In the key extractor a LFSR with same structure as the PRSG is concurrently operational as shown in the figure below.

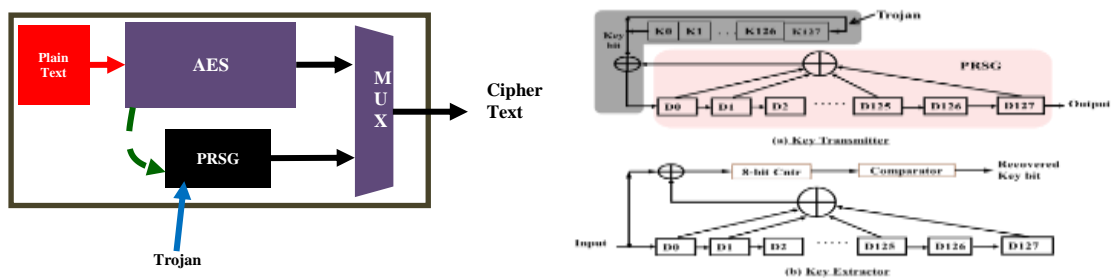

The feedback bit of this LFSR is compared (xored) with the incoming data, output of IP block, for the length of the shift register (128). If the key bit is '0' then the PRSG output at the transmitter and feedback bit of key extractor will be same and hence the output of this XOR gate will be '0' throughout and the 8-bit counter value will remain at '0', after 128 clock cycles; while, if key bit is a '1' then since the PRSG output at the transmitter is inverted, XOR gate output at the key extractor side will be '1' and the 8- bit counter value will be 128 at the end of 128 clock cycles. To cater for the channel errors, the comparator compares the counter value with 3 or 125 instead of '0' and '128'.The comparator can be set to any other values like 5 or 123, as well, without impacting design results. Whenever user data is present at the input of the IP core it sends out the encrypted output of the AES block through the output Mux gate. The output of the comparator in the malicious receiver will then be some value close to 64, as there is 50% chance of a mismatch between the feedback bit of extractor and the cipher data bits, which is deemed to be unrelated to key bits. With this approach all the 128 key bits are correctly recovered in the malicious receiver by the adversary.

**Analysis – Trojan in "Defeating HaTCh…" belongs to class $H_D$**

As the proposed HT is triggered only during the periods when there are no plain text data available at the input, it can be argued that its trigger mechanism is implicit in the valid functional behavior of the AES block, and therefore the entire AES design itself constitutes the HT trigger mechanism.

i. *Trigger Signal Dimension d(T)* This implies that its Trigger Signal Dimension (*d*) will consist of all the wires in the AES block and therefore the value of *d* will be extremely large.
ii. *Payload Propagation Delay t(T)* The HT on every occurrence of being triggered, will send out only a new single bit of the 128-bit encryption key as a payload embedded as a 128-bit vector generated by the scrambler. We can therefore argue that at a minimum the HT will need 128x128 (= 16384) clock cycles to deliver the encrypted key to the receiver (provided there is no plain text available at the input of the AES block). In general, the HT payload will be delivered over a much larger number of clock cycles as it needs to be interspersed over periods devoted to encrypted data being sent during availability of plain text at the input. Therefore, the Payload Propagation Delay (*t*) will be directly proportional and dependent on how often plain text is present at the input of the AES block – more plain text translates to larger values of *t*.
iii. *Implicit Behavior Factor α (T)* The output of the scrambler ensures that random data is sent out over the output channel during periods when no plain text is available. Therefore, our proposed HT will explicitly not manifest malicious behavior after it is triggered; thereby implying that the Implicit Behavior Factor (α) will be very high which in turn results in the HT having a very high degree of stealthiness. This implicit behavior is proven by the statistical, functional and coverage based analysis given in the paper [2].
iv. *Trigger Signal Locality l(T)* As argued above in the context of the Trigger Signal Dimension (d) parameter, as all the wires in the AES block constitute the trigger signal wires, its geographical spread will be as large as the AES block itself post physical design, and therefore, even the Trigger Signal Locality (*l*) parameter will have an extremely high value, rendering the HT harder to detect.

**Conclusion**

With the above arguments, we can firmly conclude that our proposed trojan design in [2] definitely belongs to the class $H_D$ and based on the arguments already presented in it ([2]), that it will not be detected using the HaTCh algorithm.